\documentclass[12pt]{iopart}
\usepackage{graphicx}
\usepackage{color}
\begin{document}
\title[Quantum discord and its geometric measure with
death entanglement ...] {Quantum discord and its geometric measure
with death entanglement in  correlated dephasing two qubits system}
\author{Abdel-Baset. A. Mohamed}
\address{Faculty of Science, Assiut University, Assiut, Egypt}
\begin{abstract}
The quantum correlations, including entanglement and
discord with its geometric measure, and classical correlation are
studied for  a bipartite partition of a open or closed quantum
system.
It is found that the purity of the initial state plays an important
role in the dynamics of quantum  and classical correlations.
In the dephasing model, the quantum correlations loss and  the
classical correlation gain are instantaneously  happen.
While, the purity of the initial state destroys the quantum
correlations which is resulted by the unitary interaction.
Therefore, with the purity parameter,  a particular region  in which
there is no state have quantum correlations  can be
determined.
\end{abstract}
\section{Introduction} The
development of the quantum information technology stimulates a deep
study of the properties of quantum correlations inherent in a
quantum system.
In particular, there is a problem of identification of those quantum
correlations which are responsible for the advantages of the quantum
computations in comparison with the classical ones.
Therefore, lots of interest have been devoted to the definition and
understanding of correlations in quantum systems in the last two
decades.
Specially, the definition and study of quantum and classical
correlations in quantum systems.
It is well known that the total correlation in a bipartite quantum
system can be measured by quantum mutual information \cite{h1},
which may be divided into classical and quantum parts [2-5].
The quantum part is called quantum discord(QD) which is originally
introduced by \cite{p1}.
Recently, it has been aware of the fact that quantum discord is a
more general concept to measure quantum correlation than quantum
entanglement(QE) since there is a nonzero quantum discord in some
separable mixed states \cite{p1}.
The dynamics of quantum discord and entanglement has been recently
compared under the same conditions when entanglement dynamic
undergoes a sudden death \cite{p4b,p11,p4d,ads1,ads2}. It was shown
that quantum discord presents an instantaneous disappearance at some
time points in non-Markovian regime \cite{p4b}, and exponential
decay in Markovian regime\cite{p11}.
Interestingly, it has been proven both theoretically and
experimentally that such states provide computational speedup
compared to classical states in some quantum computation
models\cite{p5,p6}.
In these contexts, quantum discord could be a new resource for
quantum computation.
\\ \indent
The calculation of quantum discord is based on numerical
maximization procedure, it does not guarantee exact results and
there are few analytical expressions including special cases
\cite{p7,p8}.
To avoid this difficulty, geometric measure of quantum discord
(GMQD) is introduced  by Ref.\cite{p9},  which measures the quantum
correlations through the minimum Hilbert-Schmidt distance between
the given state and zero discord state.
\\ \indent
Because of the unavoidable interaction between a quantum system and
its environment, understanding the dynamics of quantum and classical
correlations (CC) is an interesting line of research [14-17].
The influence of Markovian \cite{p11} and non-Markovian \cite{p12}
environment on the dynamics of QD and GMQD.
They showed that both GMQD and QD die asymptotically with
entanglement sudden death, and  the discontinuity in the decay rate
of GMQD does not always imply the discontinuity in the decay rate of
QD.
Also, they   observed that even when QD vanishes at discrete times,
GMQD disappears but not instantly.
But in our work  the dephasing environment, in which energy transfer
from the system to the environment does not occur, is considered.
Some work  has been devoted to this issue \cite{S52,S52a,S52b}, in
which, the authors show that disentanglement is dependent on the
initial condition and temperature of the environment.
This type of the environment leads to  phenomenon entanglement
sudden death, i.e.,   the entanglement can decrease to zero abruptly
and remains zero for a finite time \cite{S13,S14,S17}. Entanglement
sudden death has been experimentally observed in an implementation
using twin photons \cite{S15}, and atomic ensembles \cite{S16}.
\\ \indent
In this paper, one  considers a dephasing model, in which two qubits
embed into a multi-mode quantized field and the interaction between
the two qubits is also considered.
Therefore, the quantum correlations via quantum discord and its
geometric measure (GMQD) is compared with both quantum entanglement
and classical correlation.
\section{Measures of correlations}
\subsection{Quantum discord}
To quantify the quantum correlations of a bipartite system, no
matter whether it is separable or entangled, one can use the quantum
discord \cite{p2,p1}.
Quantum discord measures all nonclassical correlations and defined
as the difference between  total correlation and the classical
correlation with the following expression
\begin{eqnarray}
D(\rho^{AB}) &=&\mathcal{I}(\rho^{AB})-\mathcal{Q}(\rho^{AB}),
\end{eqnarray}
which quantifies the quantum correlations in $\rho^{AB}$ and can be
further distributed into entanglement and quantum dissonance
(quantum correlations excluding entanglement)\cite{p33}.
Here the total correlation between two subsystems $A$ and $B$ of a
bipartite quantum system $\rho^{AB}$ is measured by quantum mutual
information,
\begin{eqnarray}
\mathcal{I}(\rho^{AB})
&=&\mathcal{S}(\rho^{A})+\mathcal{S}(\rho^{B})-\mathcal{S}(\rho^{AB}),
\end{eqnarray}
where $\mathcal{S}(\rho^{AB})=Tr(\rho^{AB}\log\rho^{AB})$ is the Von
Neumann entropy, $\rho^{A} = Tr_{B}(\rho^{AB})$ and $\rho^{B} =
Tr_{A}(\rho^{AB})$ are the reduced density operators of the
subsystems $A$ and $B$, respectively.
The measure of classical correlation is introduced implicitly in
Ref.\cite{p1} and interpreted explicitly in the Ref.\cite{p2}.
The classical correlation between the two subsystems $A$ and $B$ is
given by
\begin{eqnarray}
\mathcal{Q}(\rho^{AB}) &=&
\max_{\{\Pi_{k}\}}[\mathcal{S}(\rho^{A})-\sum_{k}p_{k}\mathcal{S}(\rho_{k})],
\end{eqnarray}
where $\{\Pi_{k}\}$ is a complete set of projectors to measure the
subsystem $B$, and $\rho_{k}=
Tr_{B}[(I^{A}\otimes\Pi_{k})\rho_{AB}(I^{A}\otimes\Pi_{k})]/p_{k}$
is the state of the subsystem $A$ after the measurement resulting in
outcome $k$ with the probability $p_{k} =
Tr_{AB}[(I^{A}\otimes\Pi_{k})\rho^{AB}(I^{A}\otimes\Pi_{k})]$, and
$I^{A}$ denotes the identity operator for the subsystem $A$.
Here, maximizing the quantity represents the most gained information
about the system $A$ as a result of the perfect measurement
$\{\Pi_{k}\}$.
It can be shown that quantum discord is zero for states with only
classical correlations and nonzero for states with quantum
correlations.
Note that discord is not a symmetric quantity, i.e., its amount
depends on the measurement performed on the subsytem $A$ or $B$
\cite{p9}.
\subsection{Geometric measure of quantum discord}
The geometric measure of quantum discord quantifies the quantum
correlation through the nearest Hilbert-Schmidt distance between the
given state and the zero discord state \cite{p9,p10}, which is given
by
\begin{eqnarray}
D_{A}^{g} &=& \min_{\chi\in S}\|\rho^{AB}-\chi\|^{2},
\end{eqnarray}
where $S$ denotes the set of zero discord states and
$\|A\|^{2}=Tr(A^{\dagger}A)$   is the square of Hilbert-Schmidt norm
of Hermitian operators.
The subscript $A$ of $D_{A}^{g}$ implies that the measurement is
taken on the system $A$.
A state $\chi$ on $H^{A}\otimes H^{B}$ is of zero discord if and
only if it is a classical-quantum state \cite{12}, which can be
represented as
\[\chi= \sum_{k=1}^{2} p_{k}|k\rangle\langle k|\otimes\rho_{k},\]
where $\{p_{k}\}$ is a probability distribution, $|k\rangle$ is an
arbitrary orthonormal basis for $H^{A}$ and $\rho_{k}$ is a set of
arbitrary states (density operators) on $H^{B}$.
An easily computable exact expression for the geometric measure of
quantum discord is  obtained  by Ref.\cite{p9} for a two qubit
system, which can be described as follows.
Consider a two-qubit state $\rho^{AB}$ expressed in its Bloch
representation as
\begin{eqnarray}
\rho^{AB} &=&\frac{1}{4}[I^{A}\otimes
I^{B}+\sum_{i=1}x_{i}(\sigma_{i}\otimes I^{B}+I^{A}\otimes
y_{i}\sigma_{i})\nonumber\\
&&\quad\quad\qquad+\sum_{ij=1}R_{ij} \sigma_{i}\otimes \sigma_{j}],
\end{eqnarray}
where $\{\sigma_{i}\}$ are the usual  Pauli spin matrices. The
components of the local Bloch vector are
$x_{i}=Tr(\rho^{AB}(\sigma_{i}\otimes I))$ and
$y_{i}=Tr(\rho^{AB}(I\otimes\sigma_{i} ))$.
$R_{ij}=Tr(\rho^{AB}(\sigma_{i}\otimes\sigma_{j}))$ are   the
components of the  correlation matrix\cite{p9}.
Therefore, its geometric measure of quantum discord is given by
\begin{eqnarray}
  D_{A}^{g} &=&\frac{1}{4}(\|\vec{x}\|^{2}+\|y\|^{2}-k_{max}),
\end{eqnarray}
$\vec{x}=(x_{1},x_{2},x_{3})^{T}$, $R$  is the matrix with elements
$R_{ij}$ and $k_{max}$ is the largest eigenvalue of the matrix
$K=\vec{x}\vec{x}^{T}+RR^{T}$.
\subsection{Entanglement   via  Negativity}
Here, one uses the negativity\cite{S11} to measure the entanglement,
i.e., the negative eigenvalues of the partial transposition of
$\rho^{AB}$ are used to measure the entanglement of the qubits
system. Therefore, the negativity of a state $\rho^{AB}$ is defined
as
\begin{eqnarray}
N(\rho)=\max(0,-2\sum_{j}\mu_{j}),
\end{eqnarray}
where $\mu_{j}$ is the negative eigenvalue of
$(\rho^{AB}(t))^{T_{B}}$, and $T_{B}$ denotes the partial transpose
with respect to the second system.
\section{Quantum and classical correlations   in   two-qubit models }
In this section one tries to present, by examining two examples, an
physical interpretation for the relation between quantum
entanglement, quantum and classical correlations.
The first model consists of two non-interaction qubits couple with
the same quantized field under the rotating-wave approximation.
Another is the dephasing model, in which two qubits embed into a
multi-mode quantized field and the interaction between the two
qubits is also considered.
\subsection{Two non-interaction qubits couple with the field}
Here, one considers  two qubits coupled with a single-mode cavity
field, which is in the case of the  exact resonance with the qubits.
One of the pioneering potential applications on this Hamiltonian in
the context of quantum information is  '' Cooper pair
box''(qubits)\cite{11}, i.e.,  the coupled system of two Cooper pair
box (artificial atoms) and photons stored in the resonator (cavity
mode).
The cavity is sustaining non-decaying single mode field  in its
thermal state  along with the mode structure of the electromagnetic
field.
The interaction picture Hamiltonian with the rotating wave
approximation is given by
\begin{eqnarray}\label{sq1}
\hat{H}= \lambda\sum_{k=1}^{2}
(\hat{a}^{\dagger}|0\rangle_{k}\langle
1|+\hat{a}|1\rangle_{k}\langle0|),
\end{eqnarray}
where  $\hat{a}^{\dagger}$ and $\hat{a}$ are the creation and
annihilation operators for the cavity mode, $|0\rangle_{k}$ and
$|1\rangle_{k}$ denote to the ground and excited states of the
$k$-th qubit, respectively.
For the whole system (system+field), the evolution of the whole
system is characterized by the interaction between the two-qubits
system and single-mode cavity.
However from the point of the qubits, the energy transfer between
the qubits and the field happens, which is described by the
relaxation term $\hat{a}^{\dagger}|0\rangle_{k}\langle 1|$ and the
backaction term $\hat{a}|1\rangle_{k}\langle0|$.
One  assumes  that the two qubits are initially prepared in Werner
states, which is defined by
\begin{eqnarray}
\rho^{AB}(0)
&=&p|\varphi\rangle\langle\varphi|+\frac{1}{4}(1-p)\hat{I},
\end{eqnarray}
where $|\varphi\rangle=\sin\theta|11\rangle+\cos\theta|00\rangle$
and  $p$ is a real number which indicates the purity of initial
state, $\hat{I}$ is a $4\times4$ identity matrix.
But the  cavity field is initially  prepared in the vacuum state,
i.e., $\rho^{F}(0)=|0\rangle\langle0|$.
Then the initial density operation for the whole qubits-field system
is: $\rho^{ABF}(0)=\rho^{AB}(0)\otimes|0\rangle\langle0|$.
By using the above initial states, the  density matrix of the
qubits-field system with the interaction (\ref{sq1}) evolves to
$\rho(t)=\hat{U}(t)\rho^{ABF}(0)\hat{U}^{\dagger}(t)$, where the
time evolution operator $\hat{U}(t)= \exp(-i\hat{H} t)$.
The reduced density matrix, $\rho^{AB}(t)$, of two qubits  is
calculated  by tracing out the cavity field variables. Therefore,
$\rho^{AB}(t)$ is given by
\begin{eqnarray}\label{sq3}
\rho^{AB}(t)&=&
a_{1}|11\rangle\langle11|+a_{2}|00\rangle\langle00|+a_{3}(|11\rangle\langle00|\nonumber\\&&\quad
+|00\rangle\langle11|)
+a_{4}(|10\rangle\langle10|+|01\rangle\langle01|)\nonumber\\&&\quad+a_{5}(|10\rangle\langle01|+|01\rangle\langle10|)
\end{eqnarray}
with the abbreviation
\begin{eqnarray*}
a_{1}&=&\frac{1-p}{4}+\frac{p}{9}(2+\cos\varpi t)^{2}\sin^2\theta,\\
a_{2}&=&\frac{1-p}{4}+p\cos^{2}\theta+\frac{2p}{9}(1-\cos\varpi t)^{2}\sin^2\theta,\\
a_{3}&=&\frac{p}{3} (2+\cos\varpi t)\sin\theta\cos\theta,\\
a_{4}&=&\frac{1-p}{4}+\frac{p}{6}\sin^{2}\varpi t\sin^2\theta,\qquad
a_{5}=a_{4}-\frac{1-p}{4},
\end{eqnarray*}
where  $\varpi=2.4495\lambda$.
The eigenvalues of the density matrix $\rho^{AB}(t)$ are given by:$
\lambda_{1,2} = a_{4}\pm a_{5}$, and  $
\lambda_{3,4}=\frac{1}{2}[(a_{1}+a_{2})\pm\sqrt{(a_{1}-a_{2})^{2}-4a_{3}^2}].$
After some straightforward calculation,  the reduced density
matrices associated with the above states is given by
\begin{eqnarray}\label{sq4}
\rho^{A}(t)&=&\rho^{B}(t)\nonumber\\&&=
(a_{1}+a_{4})|11\rangle\langle11|+
(a_{2}+a_{4})|00\rangle\langle00|.
\end{eqnarray}
One noted that, the reduced density matrices of the qubits are
represented in  diagonal matrices. Therefore,  these  the states
$\rho^{A}(t)$ and $\rho^{B}(t)$  are classical states.
\begin{figure}
\includegraphics[width=8cm]{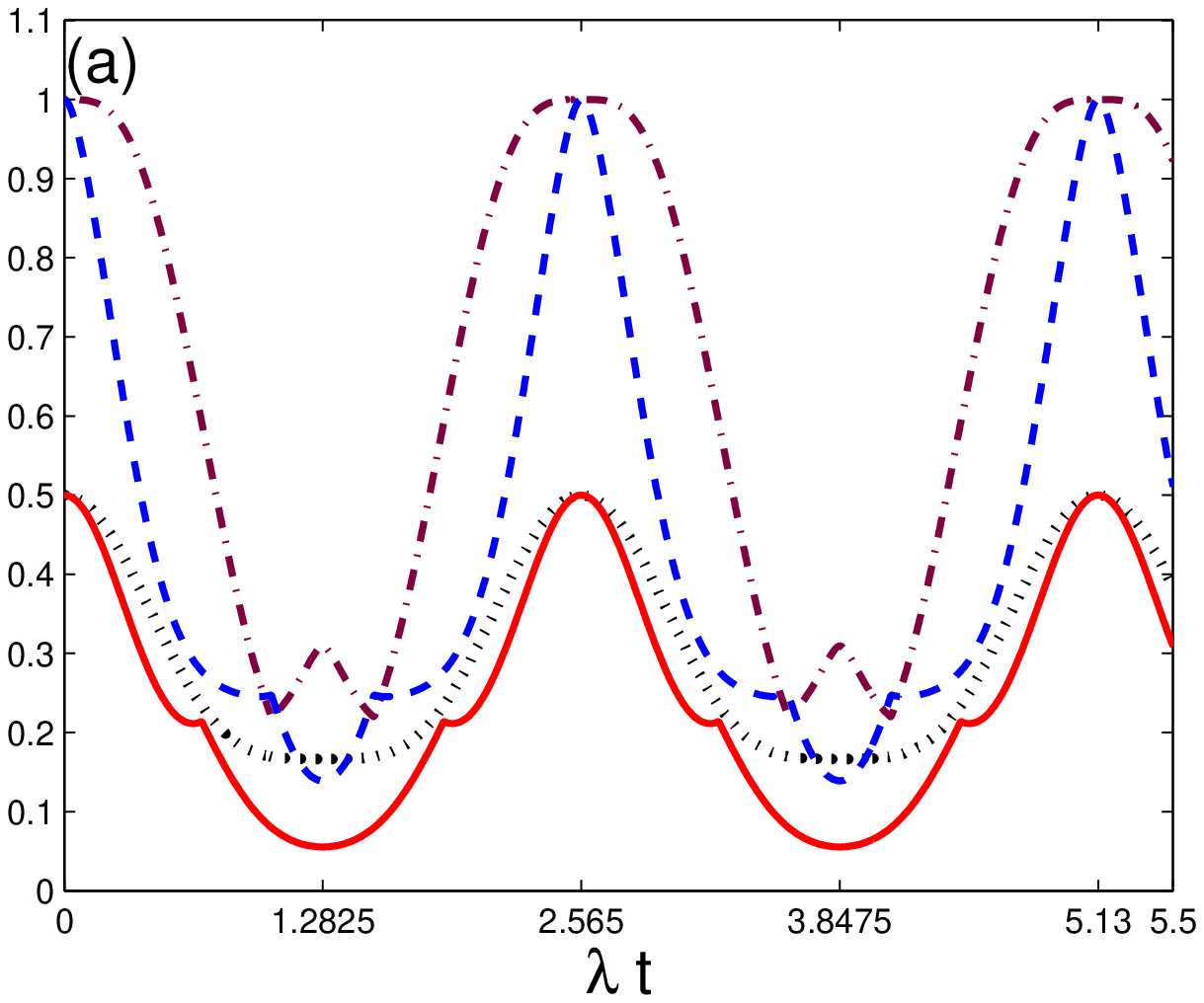}
\includegraphics[width=8cm]{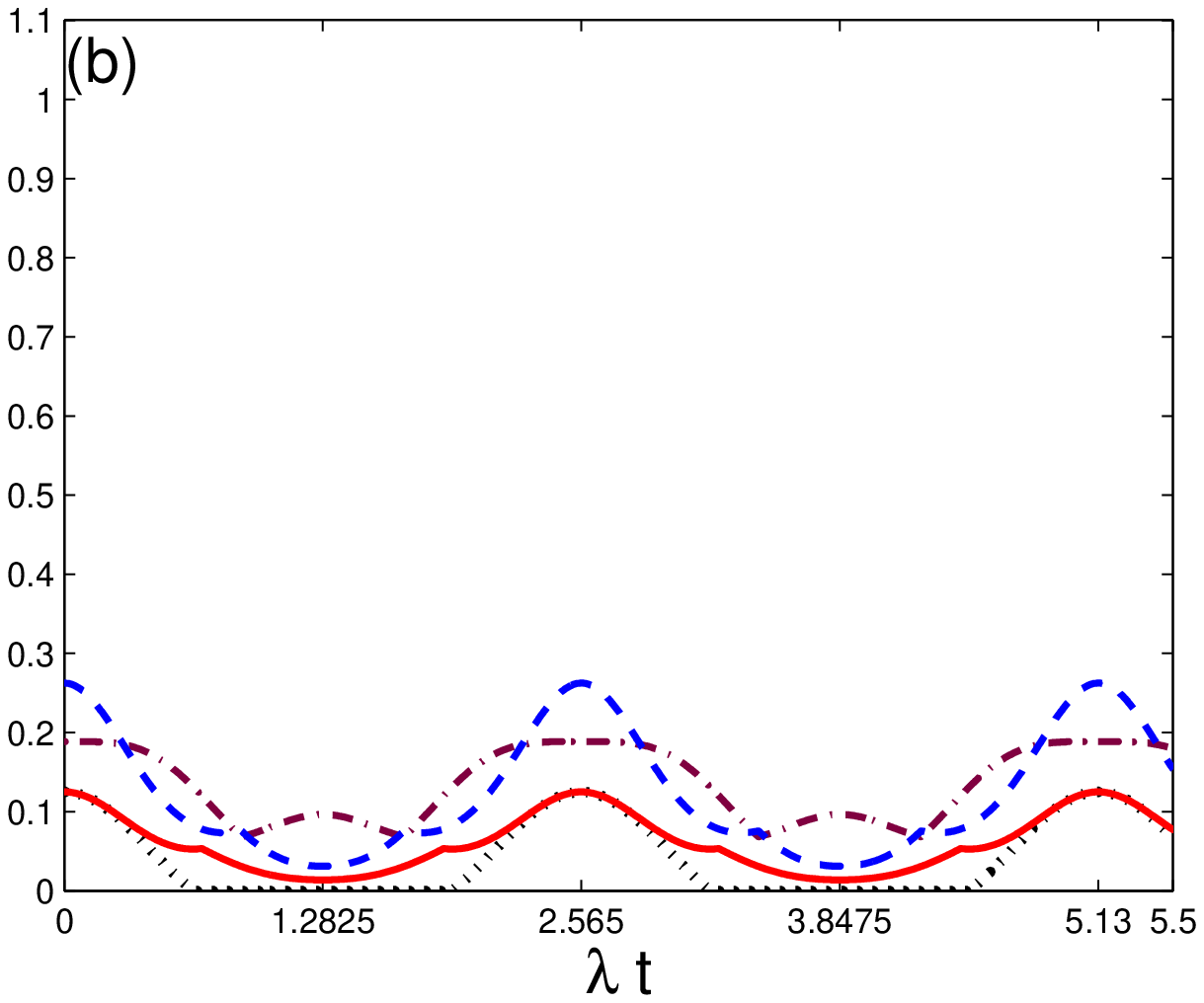}
\caption{Time evolutions of the quantum discord (dash plots), the
geometric measure of QD (sold plots), the negativity (dotted plots)
and the classical correlation (dash-dot plots)    for $p=1$ in (a)
and $p=0.5$ in (b) for $\theta=\frac{\pi}{4}$.}\label{F1}
\end{figure}
\\ \indent
For the case of the two qubits coupled with a single-mode cavity
field, the results are given in Figs.\ref{F1}a,b.
In these figures, one reports the dynamics of $D_{A}^{g}$,
$D(\rho^{AB})$, $N(\rho)$ and $Q(\rho)$  as a functions of the time
$\lambda t$ for different values of the purity of initial state
(namely $p=0.5, 1.0$) with $\theta=\frac{\pi}{4}$.
It is worth noting to mention that, because the matrix of the
initial state of the two qubits is not a diagonal matrix,  this
state  is not classical state. Therefore, its quantum correlation
have non zero value at $\lambda t=0.0$.
For $p=1.0$, these measures instantaneously oscillate and reach
their maximum (at $\lambda t=\frac{2n\pi}{\omega}, n=0,1,2,...$) and
minimum values (at $\lambda t= \frac{(2n+1)\pi}{\omega}$)  at the
same time points (see Fig.\ref{F1}a).
Because  the function $\cos\varpi t$ is a periodical function on the
scaled time with  period  $\frac{2\pi}{\varpi}$, $D_{A}^{g}$,
$D(\rho)$, $N(\rho)$ and $Q(\rho)$   evolve periodically with
respect to the scaled time with period $\frac{2\pi}{\varpi}$ (see
Figs.\ref{F1}a,b).
A rather counterintuitive feature of the QE is that it may exceed
the measures of the  GMQD and QD (see Fig.\ref{F1}a). So one can say
that GMQD and QD  are more general than QE.
\\ \indent
For $p=0.5$ (see Fig.\ref{F1}b),  one can observe that the
phenomenon of entanglement death occurs,  but this phenomenon does
not occur for GMQD and QD even when the purity $p$ is small.
Because the entanglement undergoes sudden death   while the
correlations  are  long lived, QE is not greater than QD, GMQD and
CC for some time.
Also one can see that one value of QE corresponds to many values of
QD and GMQD, meaning that the states in possession of the same
entanglement give different correlations.
This means that, there are correlations(quantum and classical) in
the intervals of entanglement death.
Therefore,  the entanglement is not the only part of quantum
correlations. This agrees with Ref.\cite{12},  which  showed that
\emph{absence of entanglement does not imply classicality}.
From Fig.1a, one can see that the intervals of vanishing negativity
disappear when $p=1$.
Therefore,   the entanglement sudden death is completely  sensitive
to the purity of the  initial-state.
\subsection{Dephasing two interaction qubits by a multimode quantized field}
\begin{figure}
\begin{center}
\includegraphics[width=8cm]{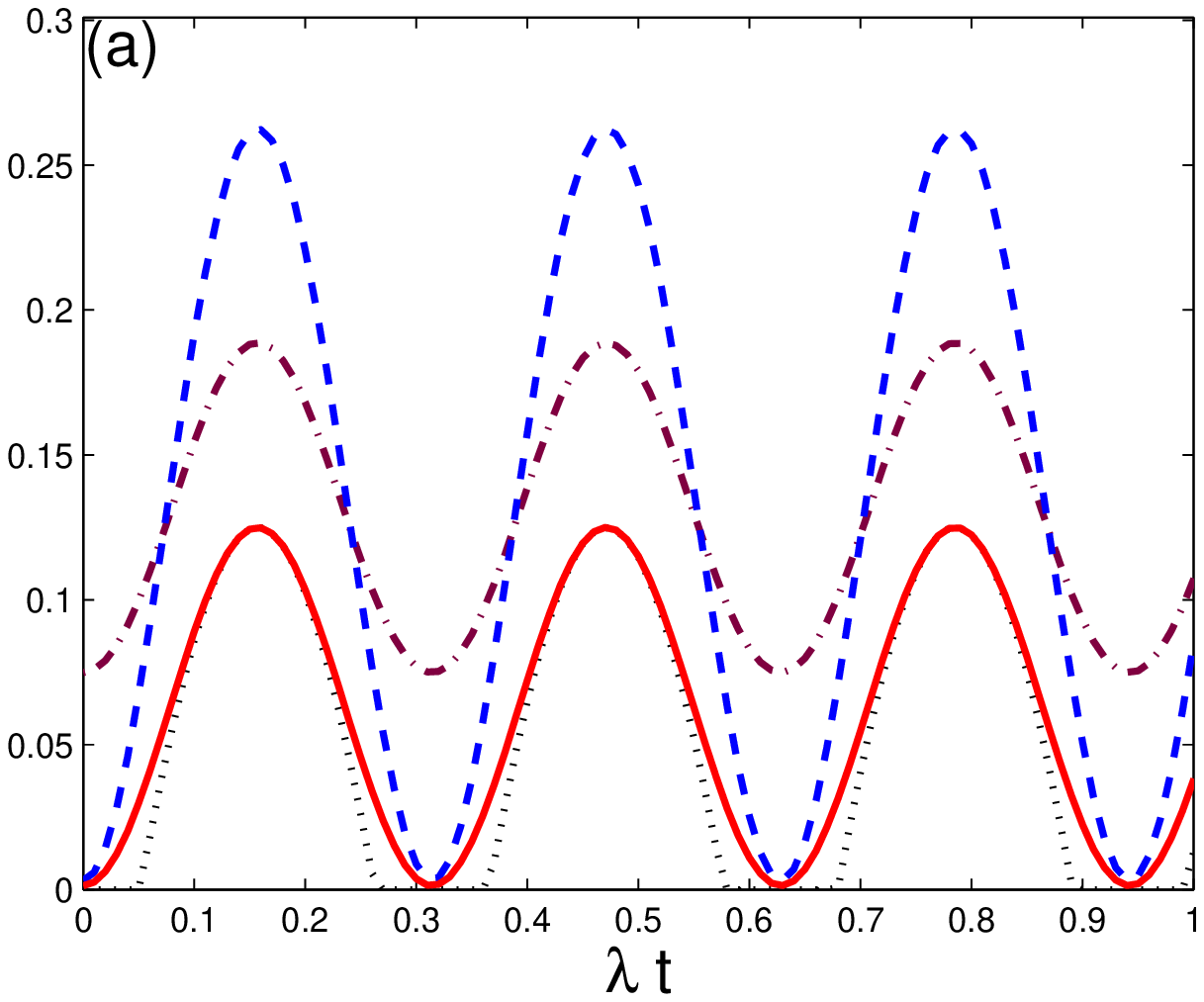}
\includegraphics[width=8cm]{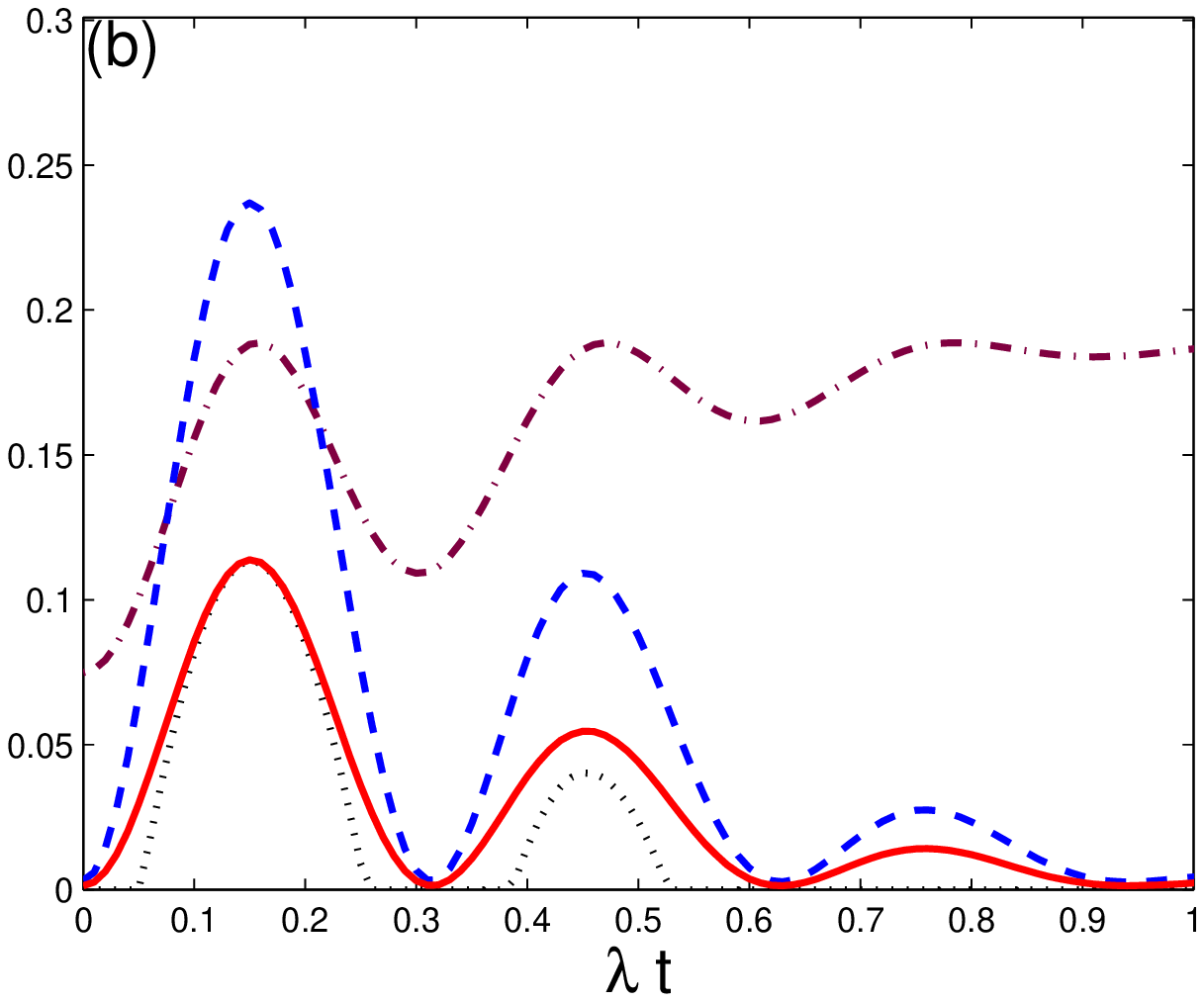}
\caption{Time evolutions   of the quantum discord (dash plots), the
geometric measure of QD (sold plots), the negativity (dotted plots)
and the classical correlation (dash-dot plots) for
$\gamma=0.0\lambda$ in (a) and $\gamma=1.0\lambda$ in (b)for
$\theta=\frac{\pi}{60}$ and $p=0.5$.}\label{F2}
\end{center}
\end{figure}
\begin{figure}
\begin{center}
\includegraphics[width=8cm]{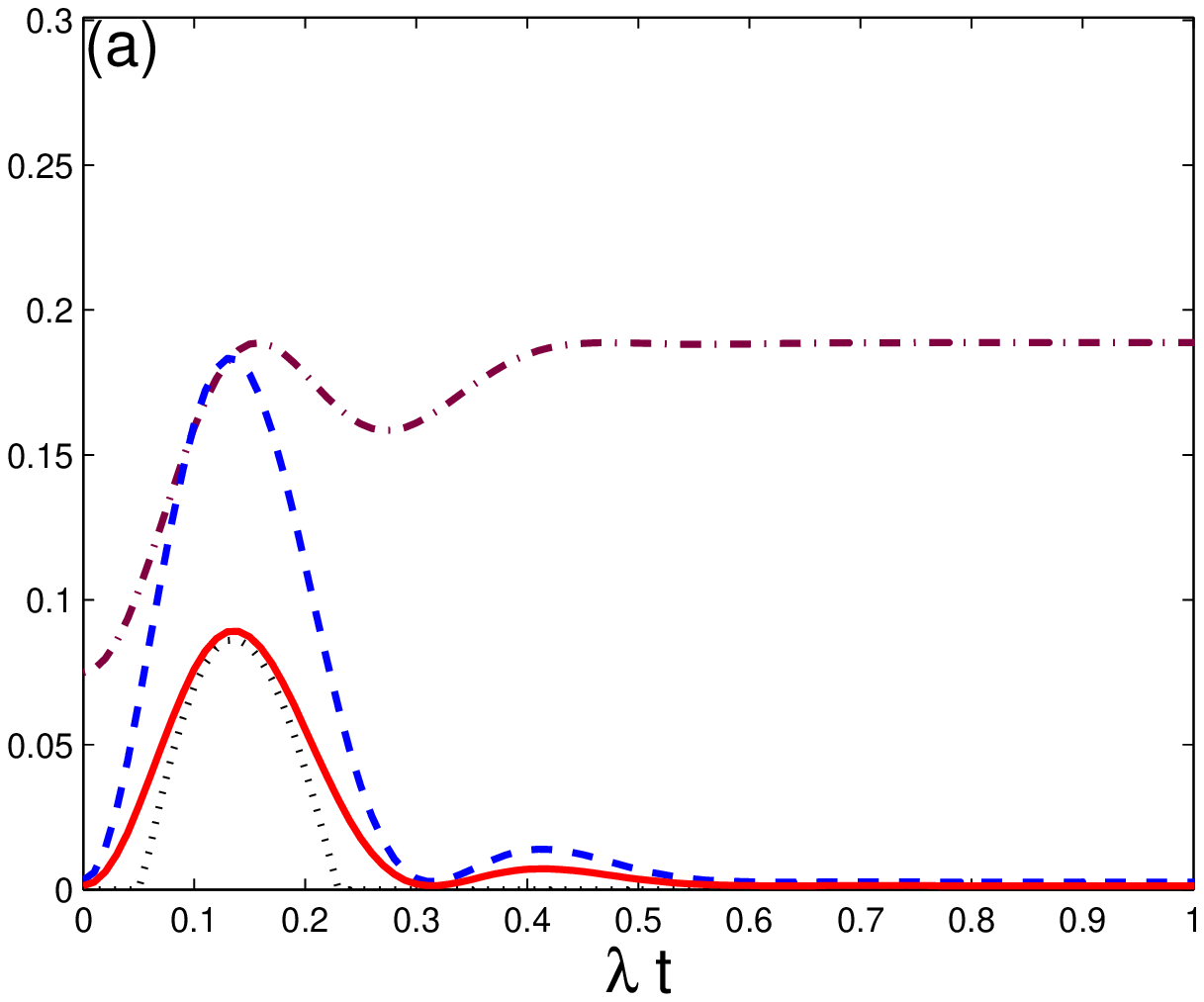}
\includegraphics[width=8cm]{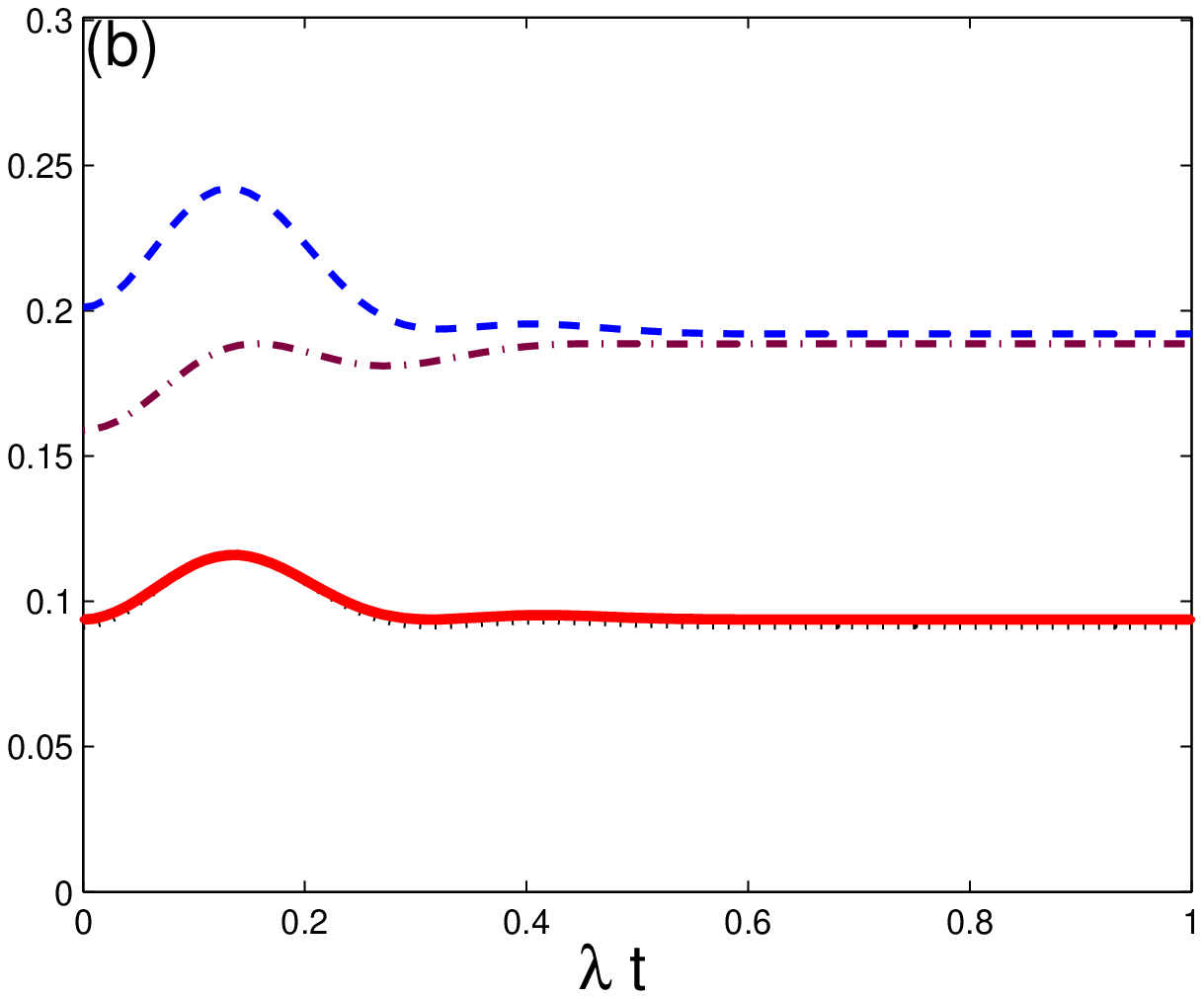}
\caption{ The same as in Fig.\ref{F2} but   for
$(\frac{\gamma}{\lambda},\theta)=(2,\frac{\pi}{60})$ in (a) and
$(\frac{\gamma}{\lambda},\theta)=(2,\frac{\pi}{3})$ in
(b)}\label{F3}
\end{center}
\end{figure}
Here,  one  considers a dephasing model of   two qubits embed into a
multimode quantized field and the interaction between the two qubits
is also considered.
The dephasing channel  case is  important situation for the open
systems, in which there is no energy transfer between the system and
environment. The Hamiltonian of this case can be written as\cite{H2}
\begin{eqnarray}\label{sq5}
\hat{H}&=&\frac{\omega_{0}}{2}
(\sigma^{z}_{A}+\sigma^{z}_{B})+\lambda
(\sigma^{+}_{A}\sigma^{-}_{B}+\sigma^{+}_{B}\sigma^{-}_{A})\nonumber\\
&&\quad+\sum_{k}\omega_{k}\hat{b}_{k}^{\dagger}\hat{b}_{k}
+\gamma_{k}(\sigma^{+}_{A}\sigma^{-}_{B}+\sigma^{+}_{B}\sigma^{-}_{A})(\hat{b}_{k}^{\dagger}+\hat{b}_{k}),
\end{eqnarray}
where $\omega_{o}$ is the qubit transition frequency and
$\omega_{k}$ the  frequency of the $k$-th field,  the   coupling
constant between the two qubits is $\lambda$ but $\gamma_{k}$ is
system-reservoir coupling constant  (dephasing channel parameter).
\\ \indent
By using the same previous  initial states but with
$|\varphi\rangle=\sin\theta|10\rangle+\cos\theta|01\rangle$, the
density matrix of the dephasing qubits-field system is given by:
$\hat{U}(t)\rho^{ABF}(0)\hat{U}^{\dagger}(t)$.
Therefore, the reduced density matrix of the two qubits  is given by
\begin{eqnarray}\label{Dph3}
\rho^{AB}(t)&=&\alpha_{1}(|11\rangle\langle11|+|00\rangle\langle00|)
+\alpha_{2}|10\rangle\langle10|\nonumber
\\&&+\alpha_{3}|01\rangle\langle01|+ \alpha_{4}|10\rangle\langle01|+\alpha_{4}^{*}|01\rangle\langle10|,
\end{eqnarray}
with the abbreviation
\begin{eqnarray*}
\alpha_{1}&=&(1-p)/4,\qquad \quad\; \;\alpha_{2}=(1+p)/4-\beta,\\
\alpha_{3}&=&(1+p)/4+\beta,\quad\quad
\alpha_{4}=\frac{p}{2}\sin2\theta-i\beta,\\
\beta&=&\frac{p}{2}  L_{d}\cos2\theta \cos2\lambda t,\nonumber\\
L_{d}&=&
\exp[-4\sum_{k}(\frac{\gamma_{k}}{\omega_{k}})^{2}(1-\cos\omega_{k})].
\end{eqnarray*}
Where $L_{d}$ is  the decoherence factor which leads to damping of
the off-diagonal terms.
One can get the eigenvalues of the density matrix $\rho(t)$ as
\begin{eqnarray}\label{sq5}
\lambda_{1,2} &=& \alpha_{1}, \quad
\lambda_{3,4}=\frac{1}{4}[1+p\pm\sqrt{(1+p)^{2}-16|\alpha_{4}|^{2}}\,]
\end{eqnarray}
The reduced density matrices associated with the above states are
given by
\begin{eqnarray}\label{sq4}
\rho^{A}(t)&=&(\alpha_{1}+\alpha_{2})|11\rangle\langle11|
+(\alpha_{1}+\alpha_{3})|00\rangle\langle00|,
\\
\rho^{B}(t)&=& (\alpha_{1}+\alpha_{3})|11\rangle\langle11|+
(\alpha_{1}+\alpha_{2})|00\rangle\langle00|.
\end{eqnarray}
The results of the case of the dephasing model   are given in
Figs.\ref{F2}-\ref{F5}.
In Fig.\ref{F2}, GMQD, QD, QE and CC are plotted as a functions of
the time $\lambda t$ for different values of system-reservoir
coupling parameter$\gamma$ (namely $\gamma/\lambda=0.0, 1,2$) with
$\theta=\frac{\pi}{60}$ and $p=0.5$.
From Fig.\ref{F2}a, one can easily find the common features of the
dynamics of GMQD, QD, QE and CC.
The previous measures of quantum and classical correlations  present
instantaneous oscillations and reach  their extreme values   at the
same time  points.
Because the entanglement undergoes sudden death for $\gamma\neq0$,
while $D_{A}^{g}$, $D(\rho)$ and $Q(\rho)$ are  long lived.
$N(\rho)$ is not greater than $D_{A}^{g}$, $D(\rho)$ and $Q(\rho)$
for small intervals  and QE attains constant values during an
intervals while  the quantum correlations (GMQD and QD) vary in
these intervals.
\\ \indent In Figs.\ref{F2}b,\ref{F3}a,b, one examines the effect of  system-reservoir
coupling  $\gamma$ on the dynamics of the previous measures of
quantum and classical correlations with $\theta=\frac{\pi}{60}$.
It is clear that  $D_{A}^{g}$, $D(\rho)$ and  $N(\rho)$ decrease
with  increasing of $\gamma$, and they have zero values.
Precisely, the dephasing parameter $\gamma$ leads to exponentially
decay for maximum values of the $D_{A}^{g}$, $D(\rho)$ and $N(\rho)$
to zero value, while $Q(\rho)$ exponentially evolves  to its
asymptotic value.
It is interesting to note that the larger the value of $\gamma$ is,
the more rapidly $D_{A}^{g}$, $D(\rho)$ and  $N(\rho)$ reach its
asymptotic values of zero.
This means that, quantum correlations, including entanglement and
discord with its geometric measure  die  asymptotically with large
values of $\gamma$.
On the contrary, the classical correlation increases  with
increasing  $\gamma$, and it have nonzero values for $\gamma>0$.
The figures show that the classical correlation   approaches an
almost steady state for large values of  $\gamma$ faster than that
for small values of  $\gamma$.
 Finally, after a very long time, the classical correlation loses it
oscillations and asymptotically reaches  its  steady state, i.e.,
the final state of the qubits reaches a classical state.
One can say that, the quantum  states with large $\gamma$ are
rapidly transformed into classical states, i.e., the processes of
quantum correlation loss and  classical correlation gain   are
instantaneously happen.
\\ \indent
To investigate the influence of $\theta=\frac{\pi}{3}$ (mixed state)
with a large value of  $\gamma=2\lambda$ on the correlations see
Fig.\ref{F3},b.
From this figure, one notes that  both  QD and CC  reach  their
asymptotic values   and  approximately  have the same behavior while
GMQD and QE have the same behavior.
This shows that the mixedness of the initial states affects on all
the previous measures   in a similar way and  it inhibits them from
going into zero.
  \indent In Figs.\ref{F4},\ref{F5},  one  examines the effect of  the
purity of the initial states on the dynamics of the  previous
measures with $\theta=\frac{\pi}{3}$ and $\gamma=0.8\lambda$.
It is clear that they decrease  with   decreasing of the  purity
parameter $p$.
When the purity $p$ is zero, all the measures  vanish, i.e, the
mixedness of the initial states have the same  effect on   all
measures.
One sees that  the influence of purity leads to: the amplitudes of
the local maxima  of   $D_{A}^{g}$, $D(\rho)$, $N(\rho)$ and
$Q(\rho)$ have exponential decay  with decreasing the parameter $p$.
When quantum correlations measures  quite vanishes,  the states
$\rho^{AB}$ finally go into a classical state  and its quantum
correlation is lost completely.
This means that, after a particular time, the purity destroys the
quantum correlations of the qubits  which is resulted by the unitary
interaction.
Therefore, in the presence of purity one can determines a particular
region, in which, there is no state have quantum correlations.
\section{Conclusions}
The  dynamics of quantum correlations, including entanglement and
discord with its geometric measure, and classical correlation in
two-qubit models are introduced for a open or closed quantum system.
It is found that the dynamics of GMQD, QD, QE and CC differ. Where,
quantum discord and its geometric measure are exist  in the region
where the entanglement is zero, which is a strong signature for the
presence of non classical correlations.
System-reservoir coupling  leads to: GMQD, QD and QE  die
asymptotically with larger system-reservoir coupling parameter.
Also, processes of quantum correlation loss and  classical
correlation gain   are instantaneously happen.
It is found that the purity of the initial states destroys the
quantum correlations by exponential decay.
Therefore, in  presence of the purity, one can determines a
particular region in which there is no state have quantum
correlations.

\end{document}